# A fourth-order finite difference scheme with accurate dispersion and adaptive dissipation for computational aeroacoustics

Yanhui Li, Yu-Xin Ren*, Youtian Su
School of Aerospace Engineering, Tsinghua University, Beijing 100084, China
*Corresponding author: ryx@tsinghua.edu.cn

## Abstract

For computational acoustics, schemes need to have low-dispersion and low-dissipation properties in order to capture the amplitude and phase of the wave correctly. To improve the spectral properties of the scheme, the authors have previously proposed a scale sensor to automatically adjust the numerical dissipation. In consequence, a fourth-order finite difference scheme with minimized dispersion and adaptive dissipation (MDAD) has been proposed [1]. In this study, we further investigate this method for the high-fidelity numerical simulation of the acoustic problems and a new dispersion control method is proposed which is different from the traditional dispersion relation preserving (DRP) approach. Firstly, the scale sensor, which quantifies the local length scale of the solution as the effective scaled wavenumber, is modified for better performance on composite waves. Then the scale sensor is applied to control both the dispersion and dissipation of the scheme. The relationships between the dispersion/dissipation parameter and the effective scaled wavenumber are analytically and artificially constructed respectively. Thus, a fourth-order finite difference scheme with accurate dispersion and adaptive dissipation (ADAD) is constructed. The approximate dispersion relation (ADR) shows that the ADAD scheme achieves accurate dispersion property at k < 2.5. The dissipation is negligible at low wave number and gradually increases after k = 1 to suppress non-physical oscillations. Several benchmark cases of computational acoustics are presented to verify the high resolution of the proposed scheme compared with the conventional spectral optimized schemes.

# 1. Introduction

Computational aeroacoustics (CAA) is an interdisciplinary field that includes fluid mechanics, acoustics, computer science, partial differential equations mathematical theory and other disciplines. The problems of CAA are usually characterized by broadband length scales, small amplitudes and large propagation distances, which places much higher demands for numerical schemes than the general computational fluid dynamics (CFD). To conduct satisfactory CAA, numerical schemes with minimal dispersion errors are desired, since the acoustic waves are nondispersive in their propagation [2,3]. As for the dissipation, the scheme with zero dissipation is generally more suitable for CAA. Whereas, it is also necessary to add artificial damping term to the scheme when the error caused by the unresolvable short waves seriously pollutes the numerical solution [4]. Due to these requirements, it is not easy to design an appropriate numerical method for CAA problems.

The dispersion relation preserving (DRP) scheme, initially proposed by Tam and Webb [3], is one of the most widely used methods in CAA. Different from the traditional central schemes with the highest possible order, the DRP scheme improves the spectral properties at the expense of the formal order of accuracy on the same stencil. This is because when the scale of the solution is relatively short compared with the grid size, these small-scale structures cannot be well resolved and the leading truncation error term is no longer a good indicator of the scheme performance [5]. However, the scheme with lower order of accuracy but better spectral properties may produce less amplitude and phase error under this circumstance. Following this idea, the DRP scheme selects difference coefficients of the stencil by minimizing an integral error of the modified wavenumber to optimize the spectral properties. As a consequence, superior dispersion and dissipation properties are achieved in both theoretical analysis (for linear advection equation) and many numerical experiments. The original fourth-order DRP scheme is also modified to compute the nonlinear acoustic pulses [6] as well as the short wave components in computational acoustics [4]. In these studies, the artificial damping term is added in the DRP scheme to suppress the non-physical oscillations emanated from regions of steep gradients and shocks.

Based on the idea of dispersion preservation, a large number of high resolution numerical schemes have been proposed by many researchers [7-12]. However, since the dispersion and dissipation properties of the numerical schemes are generally correlated, the cost functions of the spectral optimization usually contain both the dissipation and dispersion errors. Therefore, the optimized result is a compromise between the dispersion and dissipation properties. To deal with this problem, Sun et al. [13,14] proposed a sufficient condition for semi-discrete finite difference schemes to have independent dispersion and dissipation, thus the dispersion and dissipation properties can be adjusted separately without affecting each other. Based on this condition, a class of schemes with minimized dispersion and controllable dissipation (MDCD) is constructed. The MDCD schemes are characterized by the independent dispersion and dissipation properties. The dispersion is optimized by the minimization of an integral dispersion error, while the dissipation is controlled by a problem-dependent free parameter.

Recently, a class of high-order finite difference schemes with minimized dispersion and adaptive dissipation (MDAD) is constructed on the basis of the MDCD scheme [1]. The MDAD scheme has the same dispersion property as the MDCD scheme, but it releases the limitation of the MDCD scheme that the dissipation parameter needs to be manually tuned for different cases and cannot be automatically adjusted in the flow field. Actually, the numerical dissipation should be adjusted according to the scale of the flow structure. When the local scale of solution is much larger than the grid-scale, the solution is well-resolved and a small numerical dissipation is in demand. On the contrary, when the local scale of solution is close to the grid-scale, the dissipation should be sufficiently large to suppress the non-physical oscillations caused by the large dispersion error of the high wavenumber components. To realize the adaptive dissipation, Li et al. [1] devise a scale sensor which quantifies the local length scale of the solution as the effective scaled wavenumber. Then the relationship between the dissipation parameter and the effective scaled wavenumber is established with the dispersion-dissipation condition [15]. Thus, the MDAD scheme with minimized dispersion and adaptive dissipation properties is proposed.

Both the MDCD and the MDAD schemes have been successfully implemented for solving compressible flows in many cases. However, the CAA problems require much higher standards for the spectral properties of the scheme than the conventional CFD, which means the further modification is necessary. The dispersion relation preservation method [3] has been widely accepted to improve the dispersion property of the scheme for aeroacoustics problems, but the criterion for the optimal dispersion is still an open problem. For example, Weirs et al. [16] adopt the restriction in the optimization procedure that the relative dispersion error should be less than 1.5%, while the maximum relative dispersion error is chosen as 0.5% in [13]. In fact, the optimal dispersion should vary with the wavenumber components for different cases. But to our knowledge, the dispersion properties of the existing spectral-optimized schemes are predefined and cannot be adjusted, which makes a contradiction.

To address these issues, we proposed a fourth-order finite difference scheme with accurate dispersion and adaptive dissipation properties (ADAD) in this paper. The ADAD scheme is constructed based on the fourth-order MDCD scheme [13] and the scale sensor in the MDAD scheme [1]. Specifically, the scale sensor consists of the derivatives of the flow variables and is calibrated with the hormonic waves to derive the local effective scaled wavenumber of the numerical solution. In this study, we modify the scale sensor to improve the smoothness of the effective scaled wavenumber for the composite waves. Then it is applied to control both the dispersion and dissipation in the ADAD scheme. The dissipation adjustment follows the approach in the MDAD scheme, while the relationship between the dispersion parameter and the effective scaled wavenumber is analytically derived to realize accurate dispersion in a long wavenumber range (k < 2.5). This dispersion control method is different from the traditional way of dispersion preservation since it is a straightforward derivation rather than an optimization process. Therefore, the optimization criterion is no longer a limitation. Besides, the dispersion property is adaptively adjusted in the flow field for the first time, which shows better performance than the DRP scheme and the MDCD scheme in both the approximate dispersion relation (ADR) and the CAA benchmark problems.

The remainder of this paper is organized as follows. The construction of the ADAD scheme is introduced in Section 2. The spectral analysis with ADR is presented in Section 3. Several benchmark test cases are shown in Section 4 to demonstrate the performance of the proposed scheme. Conclusions are drawn in Section 5.

# 2. The fourth-order finite difference scheme with accurate dispersion and adaptive dissipation

The one-dimensional linear wave equation will be used as an example in this section to demonstrate the scheme construction procedure. That is

$$\frac{\partial u}{\partial t} + \frac{\partial f}{\partial x} = 0 \qquad (1)$$

where $f = au$ is the flux and $a$ is the constant propagation speed of waves. The semi-discrete form of Eq.(1) can be expressed as

$$\frac{\partial u_j}{\partial t} + \frac{1}{\Delta x}\left(\hat{f}_{j+1/2} - \hat{f}_{j-1/2}\right) = 0 \qquad (2)$$

on uniform grids, where $\Delta x$ is grid space and $\hat{f}_{j+1/2}$ is the numerical flux.

In general, the dispersion and dissipation properties of the numerical schemes are related, which causes contradiction in spectral optimization. To modify the dispersion and dissipation properties separately, Sun et al. [13,14] derive a sufficient condition for semi-discrete finite difference schemes to have independent dispersion and dissipation, and construct a class of finite difference schemes with low dispersion and controllable dissipation (MDCD). In this section, we will utilize the fourth-order MDCD scheme as the base scheme to construct the ADAD scheme with the scale sensor.

## 2.1 The review of the fourth-order MDCD scheme

The MDCD schemes contain $(2r+1)$ symmetrical grid points, thus the numerical flux can be written as

$$\hat{f}_{j+1/2} = \sum_{m=-r+1}^{r} b_m f_{j+m}. \qquad (3)$$

It is apparent that the highest possible order of accuracy is $2r$-th if the coefficients $b_m$ are properly selected. However, the resulting spectral properties of the scheme are not optimal under this circumstance. A high order scheme with poor spectral properties may produce unsatisfactory results when the mesh is not dense enough. Consequently, the optimized schemes usually improve the spectral properties at the expense of the formal order of accuracy. Following this idea, the MDCD schemes [13] only achieve $(2r-2)$-th order of accuracy rather than the highest possible order to obtain the decoupled spectral properties. And the decoupled dispersion and dissipation properties are controlled by the free parameter $\gamma_{disp}$ and $\gamma_{diss}$ respectively.

In this study, the MDCD scheme is specially referred to the fourth-order MDCD scheme without further notice. The numerical flux $\hat{f}_{j+1/2}$ is shown as

$$\hat{f}_{j+1/2} = \frac{1}{\Delta x}\begin{bmatrix} \left(\frac{1}{2}\gamma_{disp}+\frac{1}{2}\gamma_{diss}\right)f_{j-2} + \left(-\frac{3}{2}\gamma_{disp}-\frac{5}{2}\gamma_{diss}-\frac{1}{12}\right)f_{j-1} \\ +\left(\gamma_{disp}+5\gamma_{diss}+\frac{7}{12}\right)f_j + \left(\gamma_{disp}-5\gamma_{diss}+\frac{7}{12}\right)f_{j+1} \\ +\left(-\frac{3}{2}\gamma_{disp}+\frac{5}{2}\gamma_{diss}-\frac{1}{12}\right)f_{j+2} + \left(\frac{1}{2}\gamma_{disp}-\frac{1}{2}\gamma_{diss}\right)f_{j+3} \end{bmatrix}. \quad (4)$$

where the $\gamma_{disp}$ and $\gamma_{diss}$ are two free parameters.

The spectral properties of the scheme can be deduced with a pure harmonic wave

$$u = A(t)e^{i\omega x} \quad (5)$$

Substituting Eq.(5) into Eq.(2), we obtain the modified scale wavenumber $k'$ from

$$\frac{\partial f}{\partial x} \approx \frac{\hat{f}_{j+1/2}-\hat{f}_{j-1/2}}{\Delta x} = \frac{ik'}{\Delta x}aA(t)e^{i\omega x_j} \quad (6)$$

The modified scale wavenumber $k'$ is a function of the exact scale wavenumber $k = \omega\Delta x$. $\Re(k')$ and $\Im(k')$ are the real and imaginary part of $k'$, which indicate the dispersion and dissipation properties of the scheme respectively. The specific forms of the MDCD scheme are as follows

$$\Re(k') = \gamma_{disp}\sin 3k - \left(4\gamma_{disp}+\frac{1}{6}\right)\sin 2k + \left(5\gamma_{disp}+\frac{4}{3}\right)\sin k$$
$$\Im(k') = \gamma_{diss}\left(\cos 3k - 6\cos 2k + 15\cos k - 10\right). \quad (7)$$

It can be seen that the dispersion and dissipation properties of the MDCD scheme are independent and controlled by $\gamma_{disp}$ and $\gamma_{diss}$ respectively. Therefore, the dispersion and the dissipation properties can be optimized separately without affecting each other.

A typical strategy to improve the dispersion property is to reduce the dispersion error under a certain criterion. This method is firstly proposed in the DRP scheme [3] and has been widely used in the construction of the spectral-optimized schemes for the CAA problems. Based on this notion, Sun et al. [13] determine the optimal value of $\gamma_{disp}$ by minimizing the following integrated error function [12]

$$E = \int_0^\pi e^{-\nu k}\left(\Re(k')-k\right)^2 dk \quad (8)$$

where $e^{-\nu k}$ is the weight controlling the relative importance of errors in different wavenumber range. After comparing the results under different values of $\nu$, the optimal $\gamma_{disp}$ is set as

$$\gamma_{disp} = 0.0463783 \tag{9}$$

which is corresponding to $v=8$. For quantitative evaluation of the dispersion property, Hu et al. [2] define the well-resolved wavenumber $k_c^*$ as the maximum wavenumber satisfying $|\Re(k') - k| < 0.005$. The comparison of dispersion properties (see Fig. 1 and Fig. 2 in [1]) shows that the well-resolved wavenumber of the MDCD scheme is 32.9% larger than the fifth-order upwind linear scheme (UW5) and the sixth-order central difference scheme (C6). Whereas, the performance of the MDCD scheme in the CAA problems needs to be further investigated.

As for the dissipation, it should be non-negative in the whole wavenumber range to ensure the stability of the scheme. In the MDCD scheme, the dissipation is only controlled by the dissipation parameter $\gamma_{diss}$. The imaginary part of the modified wavenumber in Eq.(7) can be changed into the following form

$$\Im(k') = \gamma_{diss} g(\cos(k)) \tag{10}$$

where

$$g(x) = 4x^3 - 12x^2 + 12x - 4 \tag{11}$$

It is easy to prove that $g(\cos(k)) \leq 0$ is satisfied. Therefore, $\gamma_{diss} \geq 0$ will ensure the stability of the scheme when the linear advection equation has $a > 0$. And the dissipation is an increasing function of $\gamma_{diss}$. To select a proper value of $\gamma_{diss}$, the dispersion-dissipation condition proposed by Hu et al. [15] is adopted in [1] and $\gamma_{diss}$ is set as 0.012 for solving the compressible flows. But this dissipation is too large for the CAA problems to maintain the amplitude, thus the MDCD scheme with zero dissipation is utilized in this paper.

## 2.2 The modification of the scale sensor

To adjust the dissipation according to the local flow structures, a scale sensor is proposed by the authors [1], which quantifies the local scale of the solution as the effective scaled wavenumber $k_{ESW}$. The scale sensor is designed based on the physical consideration in the Taylor-series expansion at first, then the parameter in the scale sensor is calibrated with sine waves. The details can be seen in [1].

The specific form of the scale sensor for a smooth function $f$ is as follows

$$k_{ESW} = \sqrt{\frac{|\Delta f_3| + |\Delta f_4|}{|\Delta f_1| + |\Delta f_2| + \varepsilon}} \tag{12}$$

where $\Delta f_p = \frac{\partial^p f}{\partial x^p} \Delta x^p$ is numerical evaluated on the uniform mesh and $\varepsilon$ is a small positive parameter to avoid possible division by zero. A changing $\varepsilon$ is adopted in [1] to eliminates the influence of solutions with small amplitude for solving compressible flows, but in this paper $\varepsilon$ is fixed as $10^{-8}$ since the small amplitude waves need to be considered in CAA problems. When $f$ is a pure sine wave

$$f = A\sin(\omega x + \varphi) \tag{13}$$

Substituting Eq.(13) into Eq.(12), we obtain

$$k_{ESW} = \sqrt{\frac{|A\omega^3 \cos(\omega x + \varphi)\Delta x^3| + |A\omega^4 \sin(\omega x + \varphi)\Delta x^4|}{|A\omega \cos(\omega x + \varphi)\Delta x| + |A\omega^2 \sin(\omega x + \varphi)\Delta x^2| + \varepsilon}} = \omega \Delta x = k \tag{14}$$

As a result, the scale sensor derives the accurate scaled wavenumber for the harmonic waves, while for the general solution, it is a reasonable measure of the local length scale.

In the practical numerical simulation, the derivatives in Eq. (12) is discretely evaluated. Since we apply the scale sensor to control the dispersion and dissipation of the MDCD scheme, the same symmetric six-point stencil as that of the numerical flux $\hat{f}_{j+1/2}$ is adopted to calculate $k_{ESW}$ for the compactness of the implementation. However, it is found in [1] that there is a deviation between $k_{ESW}$ and the exact wavenumber for the pure sine wave, caused by the discretization errors of the derivatives. Besides, the performance of $k_{ESW}$ is not optimal when $|\Delta f_i|$ in Eq.(12) is approximated to the highest possible order of accuracy on the six-point stencil. Therefore, the evaluation of $|\Delta f_i|$ is modified with the lower order of accuracy to improve the precision of $k_{ESW}$. The detailed procedure is omitted in [1] for brevity. Thus, we explain it in this paper as a supplement.

Actually, $k_{ESW}$ shown in Eq.(12) is a combination of $k_{ESW}^{(1)} = \sqrt{\frac{|\Delta f_3|}{|\Delta f_1|}}$ and $k_{ESW}^{(2)} = \sqrt{\frac{|\Delta f_4|}{|\Delta f_2|}}$. By this way, the singularity near the critical points and the inflection points is avoided. For a pure sine wave as Eq.(13),

$$k_{ESW}^{(1)} = k_{ESW}^{(2)} = k \tag{15}$$

is analytically satisfied, thus Eq.(12) also derives the accurate scaled wavenumber $k$. However, due to the discretization errors, $k_{ESW}^{(1)} = k_{ESW}^{(2)}$ is no longer satisfied in the practical calculation if all derivates are approximated to the highest possible order of accuracy on the six-point stencil as

$$\Delta f_{1,j+1/2} = -\frac{3}{640}f_{j-2} + \frac{25}{384}f_{j-1} - \frac{75}{64}f_j + \frac{75}{64}f_{j+1} - \frac{25}{384}f_{j+2} + \frac{3}{640}f_{j+3}$$

$$\Delta f_{2,j+1/2} = -\frac{5}{48}f_{j-2} + \frac{13}{16}f_{j-1} - \frac{17}{24}f_j - \frac{17}{24}f_{j+1} + \frac{13}{16}f_{j+2} - \frac{5}{48}f_{j+3}$$

$$\Delta f_{3,j+1/2} = \frac{1}{8}f_{j-2} - \frac{13}{8}f_{j-1} + \frac{17}{4}f_j - \frac{17}{4}f_{j+1} + \frac{13}{8}f_{j+2} - \frac{1}{8}f_{j+3}$$

$$\Delta f_{4,j+1/2} = \frac{1}{2}f_{j-2} - \frac{3}{2}f_{j-1} + f_j + f_{j+1} - \frac{3}{2}f_{j+2} + \frac{1}{2}f_{j+3}$$

(16)

Therefore, the combination leads to the undesirable oscillations of $k_{ESW}$ for the harmonic waves. For instance, the numerical test of $k_{ESW}$ with Eq.(16) is presented in Fig. 1(a). This test is carried out with a pure sine wave $f = \sin(30\pi x)$ on $N=192$ uniform grids in the region $x \in [-1,1]$, and the oscillations of $k_{ESW}$ can be observed. To deal with this problem, we optimize the evaluation of derivates in the scale sensor to satisfy $k_{ESW}^{(1)} = k_{ESW}^{(2)}$. $\Delta f_3$ and $\Delta f_4$ still use the highest order form as Eq.(16), while the precision of $\Delta f_1$ and $\Delta f_2$ is reduced by two orders for further optimization. As shown in Eq.(17), $\Delta f_1$ and $\Delta f_2$ consist of three four-point sub-stencils and two free parameters $\alpha$ and $\beta$ are obtained to modify the evaluation of derivates.

$$\Delta f_{1,j+1/2} = \alpha\left(\frac{1}{24}f_{j-2} - \frac{1}{8}f_{j-1} - \frac{7}{8}f_j + \frac{23}{24}f_{j+1}\right)$$
$$+ (1-2\alpha)\left(\frac{1}{24}f_{j-1} - \frac{9}{8}f_j + \frac{9}{8}f_{j+1} - \frac{1}{24}f_{j+2}\right)$$
$$+ \alpha\left(-\frac{23}{24}f_j + \frac{7}{8}f_{j+1} + \frac{1}{8}f_{j+2} - \frac{1}{24}f_{j+3}\right)$$

$$\Delta f_{2,j+1/2} = \beta\left(-\frac{1}{2}f_{j-2} + \frac{5}{2}f_{j-1} - \frac{7}{2}f_j + \frac{3}{2}f_{j+1}\right)$$
$$+ (1-2\beta)\left(\frac{1}{2}f_{j-1} - \frac{1}{2}f_j - \frac{1}{2}f_{j+1} + \frac{1}{2}f_{j+2}\right)$$
$$+ \beta\left(\frac{3}{2}f_j - \frac{7}{2}f_{j+1} + \frac{5}{2}f_{j+2} - \frac{1}{2}f_{j+3}\right)$$

(17)

To fulfill the condition $k_{ESW}^{(1)} = k_{ESW}^{(2)}$, the cost function of the optimization is designed as

$$E = \int_0^\pi \left(k_{ESW}^{(1)} - k_{ESW}^{(2)}\right)^2 dk$$

(18)

The minimization of Eq.(18) results in

$$\alpha = \frac{1}{4}, \beta = -\frac{1}{12} \tag{19}$$

and $k_{ESW}^{(1)} = k_{ESW}^{(2)}$ is achieved in the whole wavenumber space. Therefore, Eq.(17) can be rewritten as

$$\Delta f_{1,j+1/2} = \frac{1}{96} f_{j-2} - \frac{1}{96} f_{j-1} - \frac{49}{48} f_j + \frac{49}{48} f_{j+1} + \frac{1}{96} f_{j+2} - \frac{1}{96} f_{j+3}$$
$$\Delta f_{2,j+1/2} = \frac{1}{24} f_{j-2} + \frac{3}{8} f_{j-1} - \frac{5}{12} f_j - \frac{5}{12} f_{j+1} + \frac{3}{8} f_{j+2} + \frac{1}{24} f_{j+3} \tag{20}$$

It can be seen in Fig. 1(b) that this modification of the scale sensor eliminates the oscillations of $k_{ESW}$ for the harmonic waves since $k_{ESW}^{(1)} = k_{ESW}^{(2)}$ is satisfied by the optimized forms of the derivates. To demonstrate the performance of the proposed scale sensor in the whole wavenumber range, the numerical test for $k_{ESW}$ using the sine waves $f = \sin(\omega x + \varphi)$ with different wavenumber $k$ is presented in Fig. 2. It can be noticed that after the modification, $k_{ESW}^{(1)} = k_{ESW}^{(2)} = k_{ESW}$ is obtained in the whole wavenumber range, and the smoothness of $k_{ESW}$ is improved.

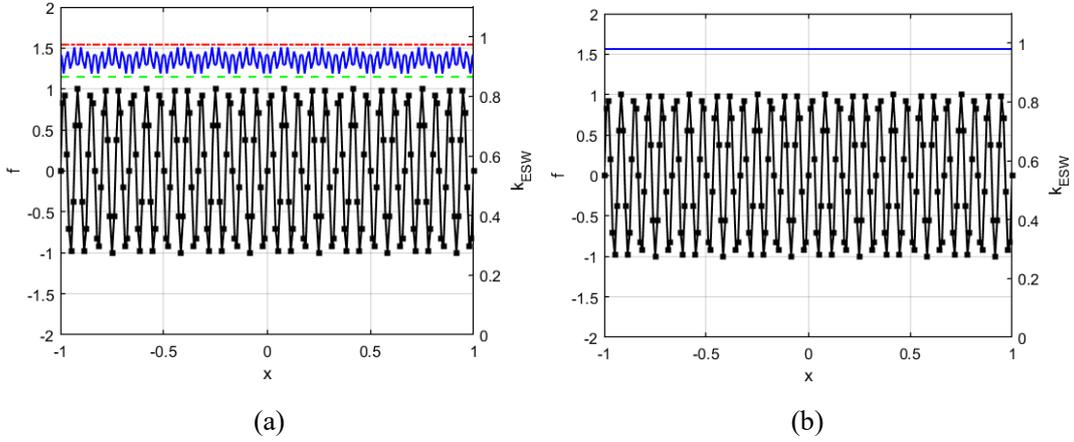

(a)              (b)

**Fig. 1.** Distributions of the scale sensor for assigned data $f = \sin(30\pi x)$ with 192 uniform grids. (a) the result of the scale sensor with the highest order discretization of derivates; (b) the result of the scale sensor with the optimized discretization of derivates. (■) the assigned data; (-.-) the value of $k_{ESW}^{(1)}$; (---) the value of $k_{ESW}^{(2)}$; (—) the value of $k_{ESW}$.

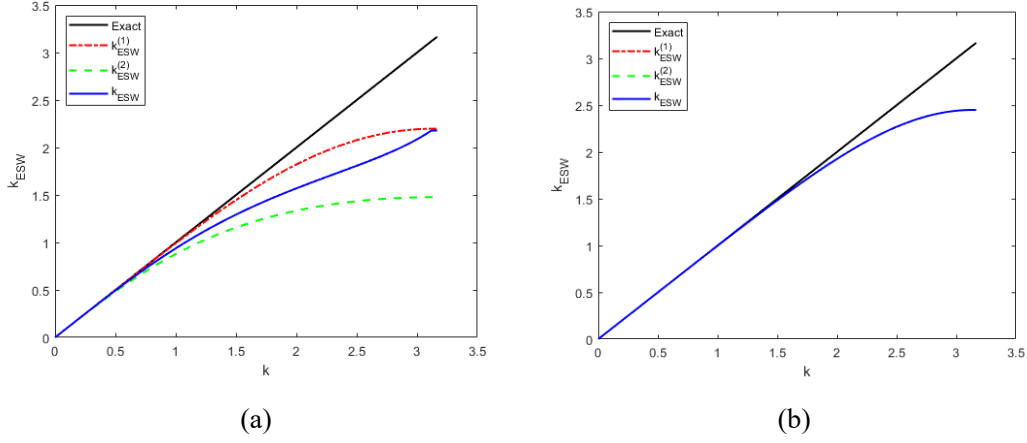

(a)                                               (b)

**Fig. 2.** The predicted effective scaled wavenumber in the wavenumber space. (a) the result of the scale sensor with the highest order discretization of derivates; (b) the result of the scale sensor with the optimized discretization of derivates.

The above optimization procedure has already been applied in [1] to improve the performance of the scale sensor for pure harmonic waves. In this paper, we further modify the form of the scale sensor to improve the smoothness of $k_{ESW}$ on composite waves. It can be seen in Eq.(14) that $\Delta f_3$ and $\Delta f_4$ differ by a factor of $k$ in magnitude. $\Delta f_1$ and $\Delta f_2$ also satisfy this relationship. Therefore, to balance the magnitude on the numerator and denominator in Eq.(12), the scale sensor is adjusted to the following form

$$k_{ESW} = \sqrt{\frac{k^*|\Delta f_3| + |\Delta f_4|}{k^*|\Delta f_1| + |\Delta f_2| + \varepsilon}} \qquad (21)$$

In the practical simulation, we firstly predict the wavenumber $k^*$ with Eq.(12) for the magnitude correction, then the effective scaled wavenumber $k_{ESW}$ is evaluated with Eq.(21). To verify the effectiveness of this modification, the numerical test carried out with a composite wave $f = \sin(2\pi e^{x+1} x)$ on $N = 192$ uniform grids in $x \in [-1,1]$ is presented in Fig. 3. It is obvious that the scale sensor with Eq.(21) obtains more stable results in this case.

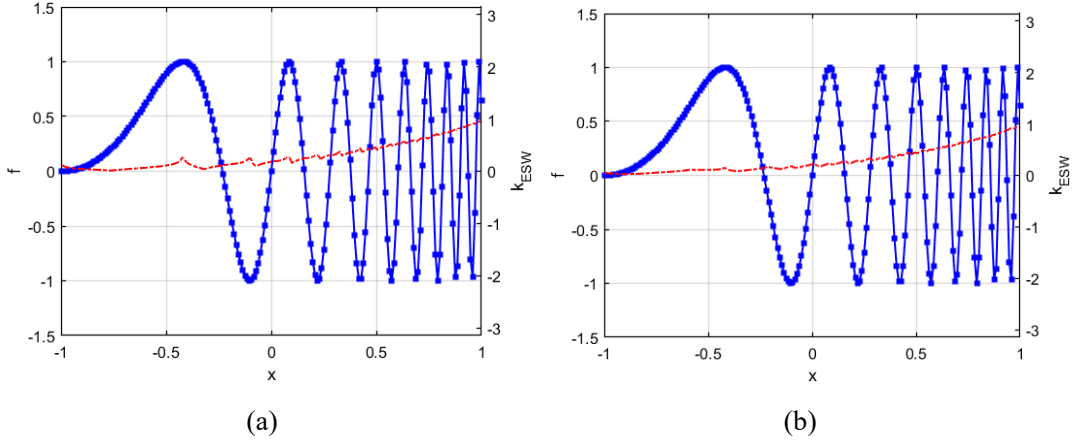

(a)  (b)

**Fig. 3.** Distributions of the scale sensor for assigned data $f = \sin(2\pi e^{x+1} x)$ with 192 uniform grids. (a) the result of the scale sensor with the highest order discretization of derivates; (b) the result of the scale sensor with the optimized discretization of derivates. (■) the assigned data; (---) the value of $k_{ESW}$.

For the pure harmonic waves, we have $k^* = k_{ESW}$ since $\sqrt{\frac{|\Delta f_3|}{|\Delta f_1|}} = \sqrt{\frac{|\Delta f_4|}{|\Delta f_2|}}$ is achieved. Therefore, the above modification does not affect the result of the scale sensor in the wavenumber space as shown in Fig. 2(b). The result in Fig. 2(b) indicates that $k_{ESW}$ can accurately predict the exact wavenumber when $k < 1.5$. Whereas, $k_{ESW}$ has a deviation from the exact result in the high wavenumber range. As address in [1], since $k_{ESW}$ monotonously increases with $k$, it is still sufficient to be used for the dissipation adaptation. But in CAA problems, more accurate effective scaled wavenumber is in demand to realize elaborate control of the dispersion property. To deal with this problem, $k_{ESW}$ is mapped to the exact scaled wavenumber before controlling the spectral properties of the scheme as

$$k = g(k_{ESW}) \quad (22)$$

The mapping function $g$ is a linear segmented function which is manually calibrated with $k_{ESW}$ and the exact wavenumber. Detailed information is shown in the Appendix A.

## 2.3 The adaptive adjustment of the dispersion and dissipation

In the previous section, we have introduced the scale sensor and modify it for the composite waves. In this section, we will apply the scale sensor to adjust the dispersion and dissipation properties of the MDCD scheme. As addressed in Section 2.1, the dispersion and dissipation of the MDCD scheme are independent and controlled by the free parameter $\gamma_{disp}$ and $\gamma_{diss}$ respectively. To realize the adaptive adjustment of the dispersion and dissipation, $\gamma_{disp}$ and $\gamma_{diss}$ in the numerical flux $\hat{f}_{j+1/2}$ (see Eq.(4)) are changed dynamically with $k_{ESW}$. Therefore, the relationships between the $\gamma_{disp}/\gamma_{diss}$ and $k_{ESW}$ are needed to be established.

The conventional method of dispersion controlling is the dispersion relation preserving method [3] which optimizes a cost function of the dispersion error to improve the dispersion property. Although it has been widely accepted in CAA, there is a problem that the optimization criterion is not clear. Different studies utilized various cost functions and optimization standards [3,4,7-12]. Actually, the optimization criterion should vary with the wave components in the flow field. For cases mainly containing low wavenumber components, dispersion optimization should be focused on the low wavenumber range and vice versa. However, the existing spectral-optimized schemes generally have a predefined dispersion property, which cannot be adjusted with the flow structures for different cases.

To deal with this problem, we proposed an accurate control method of dispersion which is different from the traditional DRP approach in this paper. The dispersion property of the MDCD scheme is uncoupled with the dissipation and the analytic form is shown in Eq.(7). Since we have obtained the effective scaled wavenumber, the relationship between $\gamma_{disp}$ and $k_{ESW}$ can be exactly derived as

$$\gamma_{disp} = \frac{k + \frac{1}{6}\sin 2k - \frac{4}{3}\sin k}{\sin 3k - 4\sin 2k + 5\sin k} \tag{23}$$

where $k = g(k_{ESW})$ is the mapped effective scaled wavenumber. Eq.(23) is a monotonically increasing function of $k$ in $k \in (0, \pi)$. When $k$ tends to zero from the right side, $\gamma_{disp}$ tends to $\frac{1}{30}$. And when $k$ tends to $\pi$, $\gamma_{disp}$ is an infinite value. Therefore, we need to set a critical wavenumber $k_c$ for Eq.(23) to limit the magnitude of $\gamma_{disp}$. In this paper, we select $k_c = 2.5$ to truncate Eq.(23), thus the relationship between $\gamma_{disp}$ and $k_{ESW}$ is a piecewise function as

$$\gamma_{disp} = \begin{cases} 0.0333339 & 0 \le k < 0.01 \\ \dfrac{k + \frac{1}{6}\sin 2k - \frac{4}{3}\sin k}{\sin 3k - 4\sin 2k + 5\sin k} & 0.01 \le k < 2.5 \\ 0.1985842 & k \ge 2.5 \end{cases} \quad (24)$$

Consequently, the dispersion property of the scheme is accurate for a long wavenumber range $(k < 2.5)$, which is superior to the traditional spectral-optimized schemes on the same stencil. The distribution of $\gamma_{disp}$ in the wavenumber space is shown in Fig. 4. By this way, we realize the accurate control of the dispersion property and the dispersion parameter is adaptively adjusted with the local length scale.

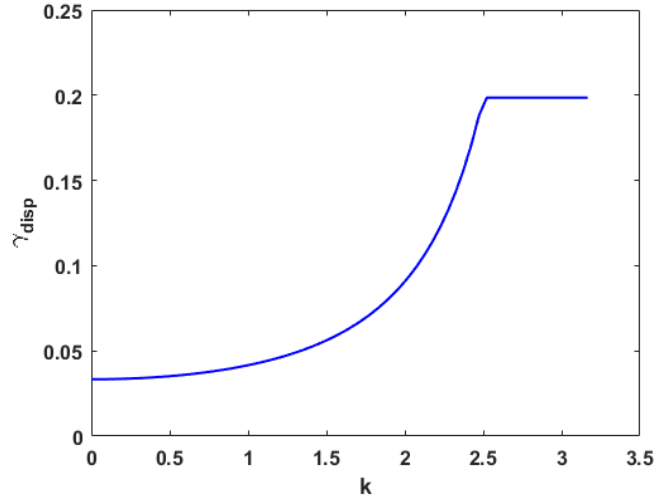

**Fig. 4.** The distribution of the dispersion parameter in the wavenumber space.

As for the adjustment of the dissipation, we adopt the similar piecewise function in [1] that

$$\gamma_{diss} = \begin{cases} 0 & 0 \le k \le 1.0 \\ 0.012\sqrt{\dfrac{k-1.0}{\pi-1.0}} & \text{otherwise} \end{cases} \quad (25)$$

In the low wavenumber range, small dissipation is already enough since the flow structures can be well-resolved by the scheme. Considering that the CAA problems are more sensitive with the dissipation, $\gamma_{diss} = 0$ is adopted for the low wavenumber components in this paper rather than $\gamma_{diss} = 0.001$ in [1]. When $k > 1$, the dissipation gradually increases to damp the non-physical high frequency oscillations. Fig. 5. presents he distribution of $\gamma_{disp}$ in the wavenumber space. It should be noticed that the relationship between $\gamma_{disp}$ and $k_{ESW}$ is artificially determined based on experience rather than analytically derived. Other approaches either empirically or based on some physical considerations can also be applied.

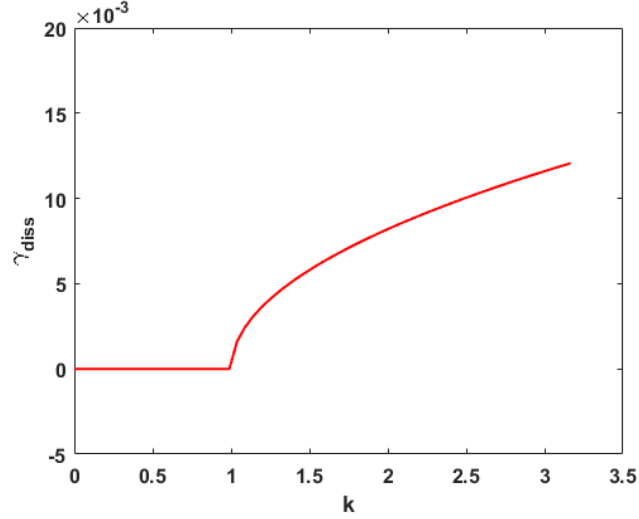

**Fig. 5.** The distribution of the dissipation parameter in the wavenumber space.

## 3.3 The spectral properties of the scheme

It has been shown in Section 2.1 that the spectral properties of the linear schemes can be analytically obtained by Fourier analysis. Nonlinear schemes, however, do not have a straightforward method to derive the analytical spectral properties. To address this problem, the approximate dispersion relation (ADR) is proposed by Pirozzoli [17]. This method numerically evaluates the modified wavenumber of the nonlinear schemes. Then the dispersion and dissipation properties of the nonlinear schemes can be analyzed. In this section, we will compare the spectral properties of the ADAD scheme with the DRP scheme [6] and the MDCD scheme [13]. The DRP scheme is a kind of central schemes with zero dissipation and the dissipation parameter of the MDCD scheme is also set to zero considering the non-dissipative feature of the CAA problems. The DRP and MDCD schemes are both linear schemes and their spectral properties can be analytically derived following the procedure in Section 2.1. Whereas, the adaptive adjustment of $\gamma_{disp}$ and $\gamma_{diss}$ introduces nonlinearity in the ADAD scheme, thus the dissipation and dispersion properties of the ADAD scheme is obtained with the ADR method.

Fig. 5 and Fig. 6 present the comparison of dispersion properties and the absolute dispersion error respectively. It is obvious that the ADAD scheme has better dispersion compared with the DRP and MDCD schemes, since it has negligible dispersion error in a long wavenumber range $(k < 2.5)$. For quantitative comparison, we also exam the maximum wavenumber $k_c^*$ under the tolerance of the dispersion error $|\Re(k') - k| < 0.005$ [2]. $k_c^*$ of the DRP, MDCD and ADAD schemes are 1.169, 1.296, 2.500 respectively. Therefore, the ADAD scheme achieves a significantly improved dispersion property by the adaptive adjustment of $\gamma_{disp}$.

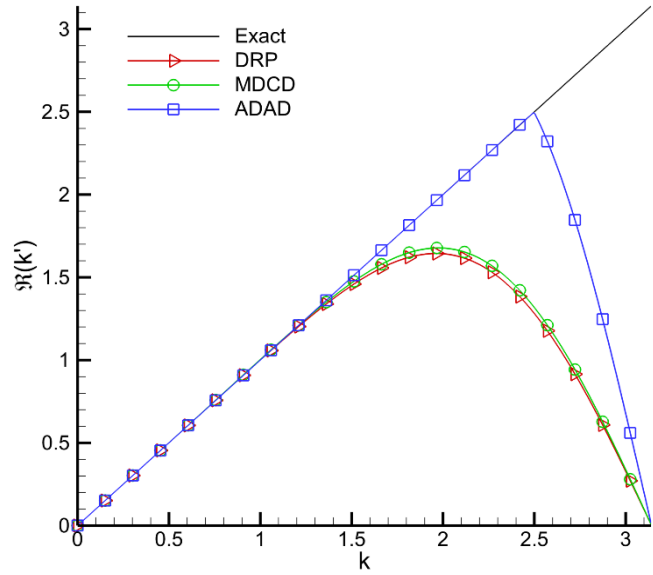

**Fig. 6.** Comparisons of dispersion properties for different schemes.

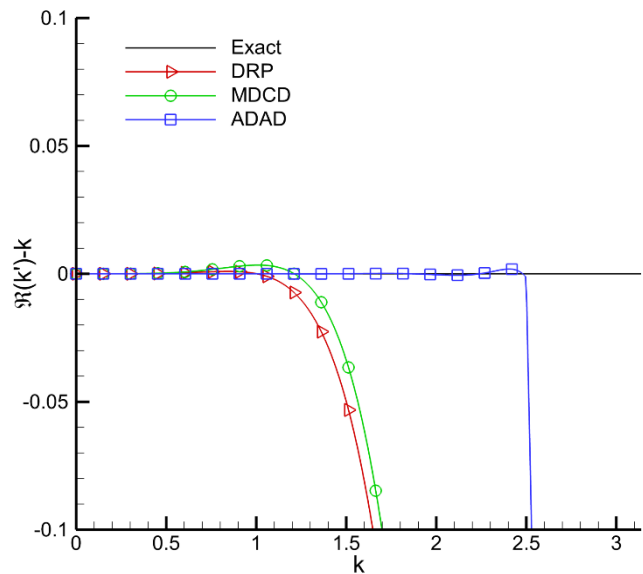

**Fig. 7.** Comparisons of absolute dispersion error for different schemes.

As for the dissipation property, both the DRP and MDCD schemes have zero dissipation in the whole wavenumber range as shown in Fig. 8. The ADAD scheme also adopts the zero dissipation in the low wavenumber range $(k<1)$. But the dissipation gradually increases according to the predefined function. This approach is to achieve high resolution for the flow structures with large scale and provide sufficient robustness for high wavenumber components meanwhile [1].

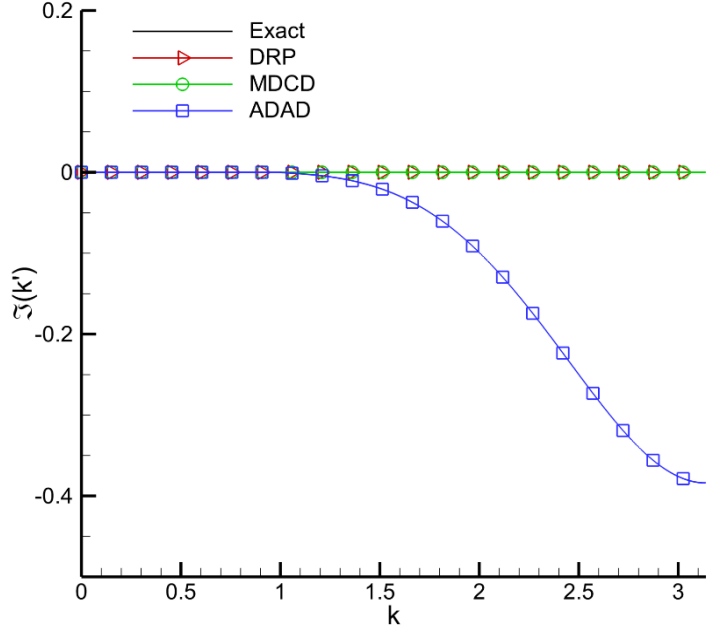

**Fig. 8.** Comparisons of dissipation properties for different schemes.

## 4. Numerical tests

Several benchmark cases in CAA are calculated to verify the performance of the proposed ADAD scheme. The results are compared with those of the DRP and MDCD schemes to confirm the high accuracy in dispersion and stableness in dissipation of the ADAD scheme. The test cases include the linear advection equation, linearized Euler equations and Euler equations in one-dimension and the linearized Euler equations in two-dimension, all of which involve the long-distance propagation of waves. The low storage fourth order Runge–Kutta scheme [18] is implemented in the time advancement. The CFL number is set as 0.3 for all cases to decrease the influence of time advancing method on the performance of spatial difference schemes. To apply the ADAD scheme in solving the Euler equations, the effective scaled wavenumber and the corresponding dispersion/dissipation parameter are evaluated on each grid interface with the split characteristic fluxes before the start of a new time-step [1]. This strategy is to improve the computing efficiency and accuracy of the dispersion/dissipation adjustment, since the reconstruction of the numerical flux is based on the split characteristic fluxes.

## 4.1 One-dimensional linear advection equation

This case [13] is adopted to check the dispersion and dissipation of the schemes under the one-dimensional linear condition. The governing equation is

$$\begin{cases} \dfrac{\partial u}{\partial t} + \dfrac{\partial u}{\partial x} = 0 \\ u(x,0) = u_0(x) \end{cases}$$

with the periodic boundary condition. The initial condition is set as

$$u_0(x) = \frac{1}{m}\sum_{k=1}^{m}\sin(2\pi k x)$$

And the analytic solution can be easily obtained by the characteristic method as

$$u(t,x) = \frac{1}{m}\sum_{k=1}^{m}\sin(2\pi k(x-t))$$

This case is characterized by the wave packet with different length scales. For larger $m$, higher wavenumber components are involved, thus this case can be used to exam the performance of the schemes for composite waves. The solution is advanced in time up to $t=1.0$ with different grid points on the computational domain $x \in [0,1]$.

Fig. 9 illustrates the L2 error of the numerical solution with different wave elements $m=5,10,15,20$. It can be seen that the ADAD scheme produces much lower error in most situations compared with the DRP and MDCD schemes. Besides, both the DRP and MDCD schemes achieve the theoretical fourth-order of accuracy, while the ADAD scheme exhibits approximate fifth-order of accuracy. This phenomenon is due to the adaptive adjustment of the dispersion parameter in the ADAD scheme, which significantly reduces the phase error of the solution.

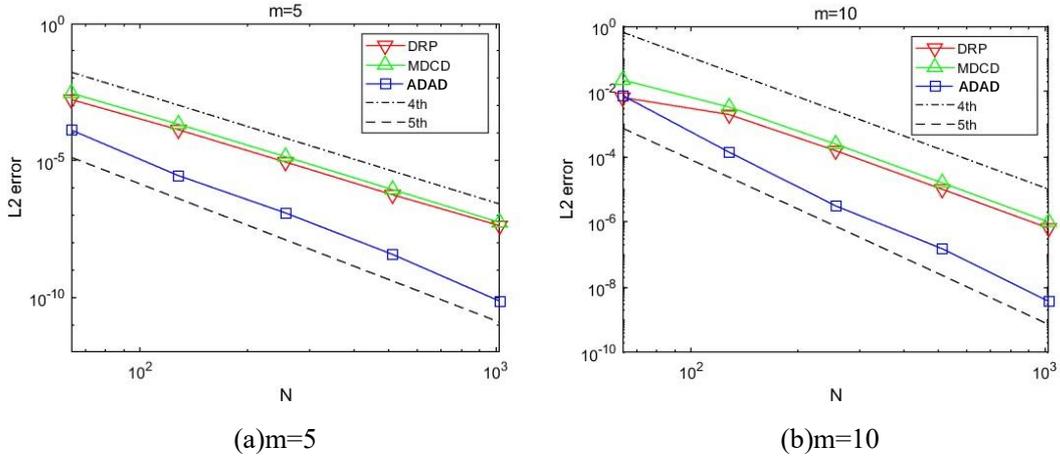

(a) m=5    (b) m=10

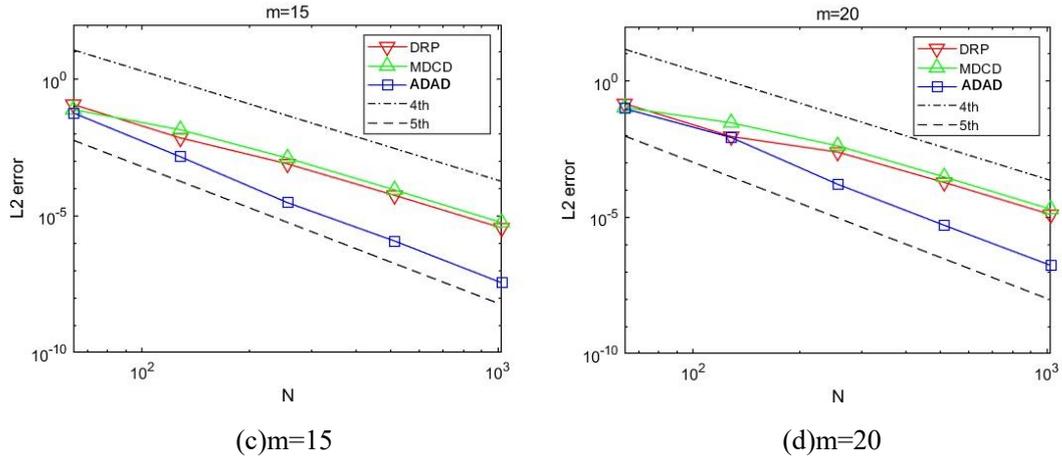

(c)m=15    (d)m=20

**Fig. 9.** The rate of convergence in terms of $L_2$ errors

In order to better demonstrate the accurate dispersion property of the ADAD scheme, the solution with $m = 20$ is presented in Fig. 10. It is obvious that only the ADAD scheme accurately captures the phase of the solution, while the other two schemes have different phase error.

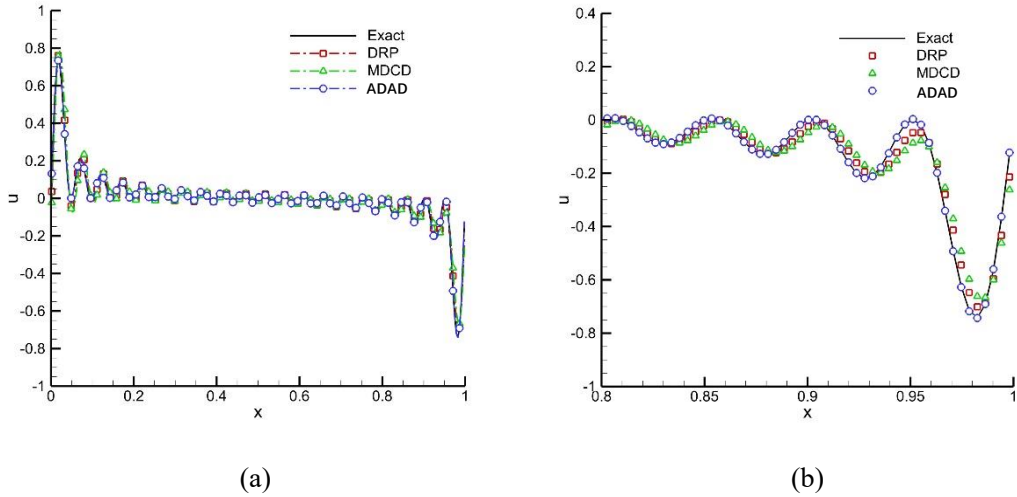

(a)    (b)

**Fig. 10.** The one dimensional linear advection problem, (a) distributions of solution at t = 1.0;

(b) the enlarged portion of (a) near the right boundary.

## 4.2 One-dimensional linearized Euler equations

This case describes two gaussian acoustic waves initially superposed in the center of the flow field and travelling to the left and right respectively. To this end, we consider the one-dimensional linearized Euler equations [19] as

$$\frac{\partial \mathbf{U}}{\partial t} + \mathbf{A}\frac{\partial \mathbf{U}}{\partial x} = 0$$

where $\mathbf{U}$ and $\mathbf{A}$ are defined by

$$\mathbf{U} = \begin{pmatrix} \rho \\ u \\ p \end{pmatrix}, \mathbf{A} = \begin{pmatrix} u_0 & \rho_0 & 0 \\ 0 & u_0 & \frac{1}{\rho_0} \\ 0 & \gamma p_0 & u_0 \end{pmatrix}$$

The original form of the linearized Euler equations can be rewritten as the characteristic form

$$\frac{\partial \mathbf{W}}{\partial t} + \mathbf{B}\frac{\partial \mathbf{W}}{\partial x} = 0$$

where $\mathbf{W}$ and $\mathbf{B}$ are defined by

$$\mathbf{W} = \begin{pmatrix} \frac{p}{\rho_0 c} - u \\ \rho - \frac{p}{c^2} \\ \frac{p}{\rho_0 c} + u \end{pmatrix}, \mathbf{B} = \begin{pmatrix} u_0 - c & 0 & 0 \\ 0 & u_0 & 0 \\ 0 & 0 & u_0 + c \end{pmatrix}$$

and $c = \sqrt{\gamma \frac{p_0}{\rho_0}}$ is the sound speed. The former equations include three waves : the acoustic waves $w_1$ and $w_3$ as well as the entropy wave $w_2$, travelling with the speed of $u_0 - c$, $u_0 + c$, $u_0$ respectively. As for the initial condition, it is set as

$$\rho = p = 0.5 \exp\left(-\ln(2)\left(\frac{x}{3}\right)^2\right), u = 0$$

and $\rho_0 = 1.0$, $p_0 = \frac{1}{\gamma}$, $u_0 = 0.1\sqrt{\frac{\gamma p_0}{\rho_0}}$. The solution is integrated in time up to $t = 100$ on an $N = 250$ mesh in the region $x \in [-200, 200]$.

The results can be seen in Fig. 11. The exact solution is analytically obtained from the characteristic equations. Oscillations can be observed in the results of all schemes, since the gauss wave profile contains broadband length scales and the dispersion error of the scheme in the high wavenumber range is inevitable. However, the ADAD scheme produces obviously lower trailing oscillations behind the acoustic waves compared with the DRP and MDCD schemes, which verify the improvement of dispersion property. Additionally, the ADAD scheme maintains comparable wave crest with the other schemes, which confirms the high resolution of the proposed scheme in CAA problems.

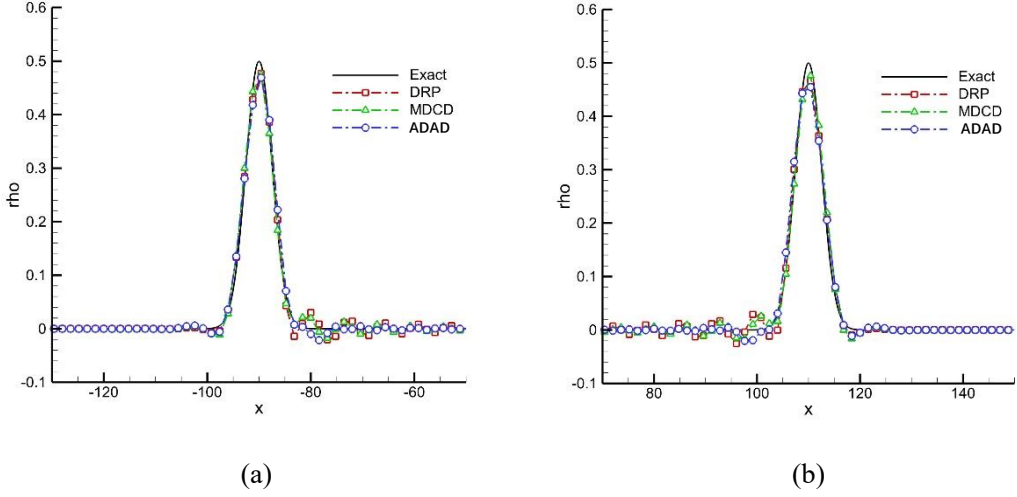

(a)                                                              (b)

**Fig. 11.** The one-dimensional linearized Euler problem at t = 100, (a) the left acoustic wave;

(b) the right acoustic wave.

## 4.3 One-dimensional Euler equations

In this case, a sound wave packet containing acoustic turbulent structures is propagated periodically in the flow field governed by the one-dimensional Euler equations [13]. This case is featured by the broadband length scales and small amplitude, thus it is suitable to evaluate the performance of numerical schemes in CAA. The initial condition is given as follows

$$p(x,0) = p_0 \left(1 + \varepsilon \sum_{k=1}^{N/2} \sqrt{E_p(k)} \sin(2\pi k(x + \psi_k))\right)$$

$$\rho(x,0) = \rho_0 \left(\frac{p(x,0)}{p_0}\right)^{1/\gamma}$$

$$u(x,0) = u_0 + \frac{2}{\gamma-1}\frac{c(x,0)}{c_0}$$

where $E_p(k) = \left(\frac{k}{k_0}\right)^4 e^{-2(k/k_0)^2}$ is the energy spectrum of the acoustic wave packet. $E_p$ reaches the maximum value when $k = k_0$. Therefore, the high wavenumber elements are more energetic with a larger value of $k_0$. In this paper, we adopt $k_0 = 12$ on a $N = 128$ grid. In addition, $\psi_k$ is a random number uniformly distributed on 0 to 1. $\varepsilon = 0.001$ is the parameter which determines the intensity of the acoustic turbulence. And $c = \sqrt{\dfrac{\gamma p}{\rho}}$ is the sound speed. The simulation is carried out in $x \in [0,1]$ for one period of time with $t = 0.917$. The periodic boundary conditions are imposed on the left and right side of the computational domain.

The numerical results in terms of density distribution are shown in Figs. 12. The exact solution is the same as the initial condition since the acoustic wave propagates periodically without dissipation and dispersion error. It can be seen that all the schemes successfully capture the flow structures. However, the enlarger portion of the small structures shows that the ADAD scheme has the best retention of phase and amplitude of the wave compared with the DRP and MDCD schemes.

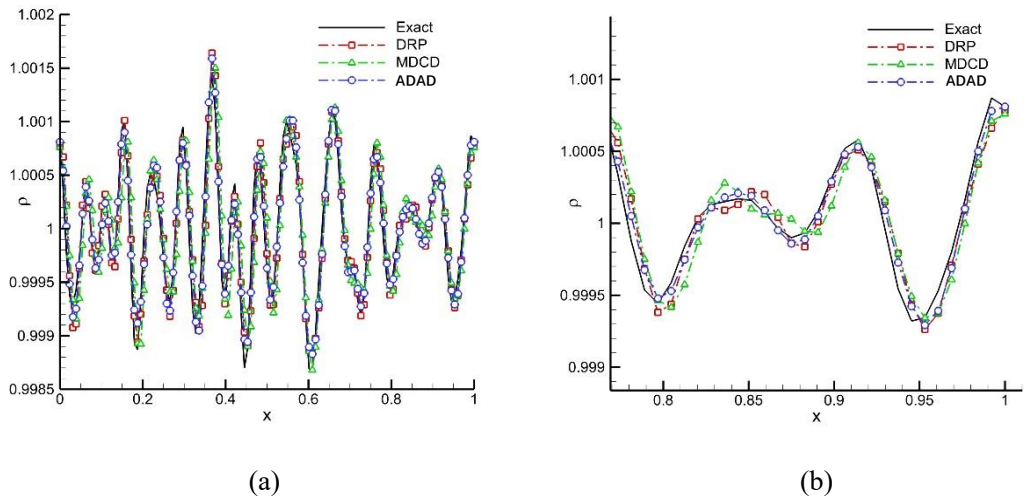

(a)              (b)

**Fig. 12.** The one-dimensional broadband wave propagation problem at t = 0.917, (a) distributions of density; (b) the enlarged portion of the small structures.

## 4.4 Two-dimensional linearized Euler equations

This case [2,20] is utilized to verify the performance of the ADAD scheme in the two-dimensional CAA problems. For this purpose, the reflection of an acoustic pulse from the solid wall at $y = 0$ is simulated, which is governed by the two-dimensional linearized Euler equations

$$\frac{\partial \mathbf{U}}{\partial t} + \frac{\partial \mathbf{E}}{\partial x} + \frac{\partial \mathbf{F}}{\partial y} = 0$$

where

$$U = \begin{pmatrix} \rho \\ u \\ v \\ p \end{pmatrix}, E = \begin{pmatrix} M_x\rho + u \\ M_x u + p \\ M_x v \\ M_x p + u \end{pmatrix}, F = \begin{pmatrix} M_y\rho + v \\ M_y u \\ M_y v + p \\ M_y p + v \end{pmatrix}$$

$M_x$ and $M_y$ are the Mach numbers of the mean flow in the x- and y-directions. In this case, we set them as $M_x = M_y = 0$ to obtain a stationary mean flow. The computational domain is $(x, y) \in [-400, 400] \times [0, 400]$ and the initial condition is given by

$$\rho = p = \exp\left(-\ln 2 \frac{x^2 + (y-25)^2}{9}\right)$$
$$u = v = 0$$

which indicates an acoustic pulse initially placed at $(0, 25)$. The bottom boundary is considered as a slip wall and the other sides adopt the outflow boundary condition. The final simulation time is $t = 300$. At this time, the sound wave motivated by the acoustic pulse has already reflected by the solid wall and propagated for a relatively long distance.

Fig. 13 illustrates the density contours obtained by the DRP, MDCD and ADAD schemes on a uniform $600 \times 300$ mesh. The propagation and the reflection of the sound wave can be observed. For quantitative comparison of the results, profiles of the density along the diagonal line $y = x + 200$ are shown in Fig. 14. The result of the ADAD scheme has less oscillations compared with that of the DRP and MDCD scheme, due to the improvement in spectral properties.

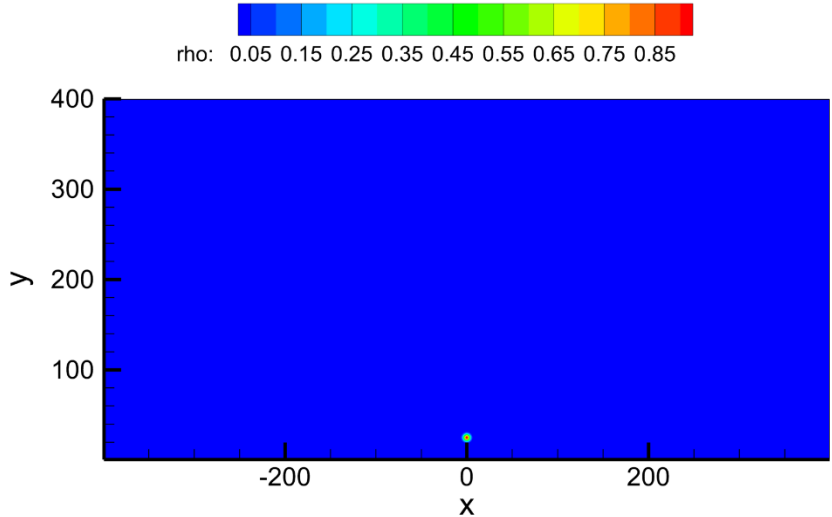

(a) t=0

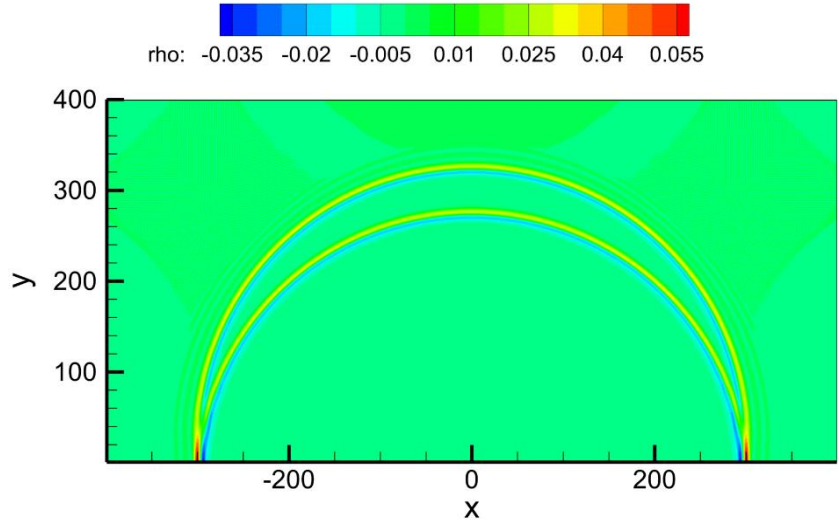

(b)t=300

**Fig. 13.** Density contours for the two-dimensional sound wave reflection problem.

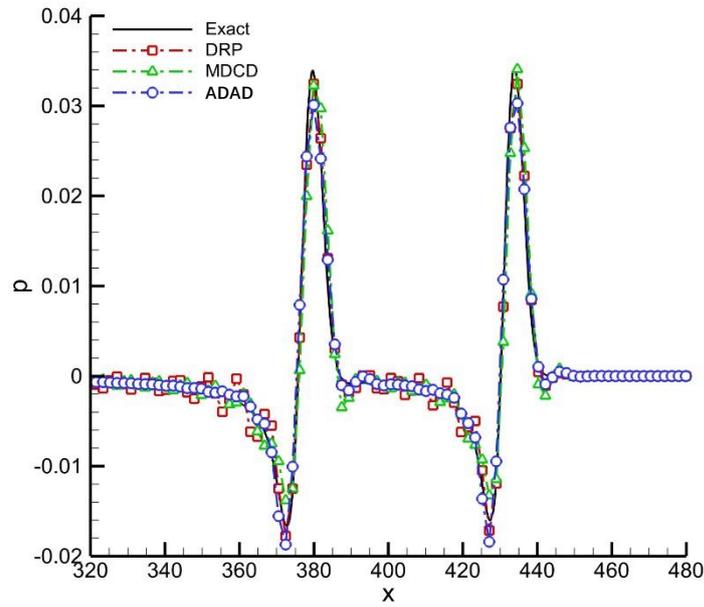

**Fig. 14.** Density contours for the two-dimensional sound wave reflection problem.

## 5. Conclusions

In this study, a fourth-order finite difference scheme with accurate dispersion and adaptive dissipation (ADAD) is proposed for solving the acoustic problems. Different from the conventional DRP scheme which has been widely applied in CAA, the ADAD scheme directly controls the numerical dispersion and dissipation by the scale sensor [1] rather than utilizing an optimization procedure. To construct the ADAD scheme, the scale sensor which is firstly proposed by authors in [1] is further modified to device more smooth result on the composite waves. Specifically, the original scale sensor is firstly introduced in Section 2.2, which consists of the derivatives of the solution and is calibrated to quantify the local scale of the solution in the form of the effective scaled wavenumber. Then a correction step is involved to balance the magnitudes on the numerator and denominator of the original form. The numerical test of the new scale sensor shows that this modification improves the performance on the composite waves. To implement the scale sensor in controlling the spectral properties, the relationship between the dispersion parameter and the effective scaled wavenumber is directly derived based on the analytical form of the dispersion property. As for the dissipation adjustment, we adopt a similar strategy as in [1]. The approximate spectral properties obtained by ADR [27] verify that the ADAD scheme achieve accurate dispersion in a long wavenumber range ($k < 2.5$). The dissipation maintains zero in the low wavenumber range ($k < 1$) and gradually increases to damp the high frequency non-physical oscillations. To confirm the performance of the proposed scheme, a variety of benchmark cases in CAA are calculated. The results show that the ADAD scheme provides better phase maintenance and comparable amplitude of the sound waves compared with the DRP and MDCD schemes.